\documentclass[aps,prb,twocolumn,showpacs]{revtex4}
\usepackage{bm}

\usepackage{amsmath}
\usepackage{amssymb}
\usepackage{graphicx}
\usepackage{psfrag}

\newcommand{\fr}{\frac}
\newcommand{\sq}{\sqrt}
\newcommand{\lbl}{\label}
\newcommand{\mm}{\!-\!}

\newcommand{\beq}{\begin{equation}}
\newcommand{\eeq}{\end{equation}}
\newcommand{\beqar}{\begin{eqnarray*}}
\newcommand{\eeqar}{\end{eqnarray*}}

\newcommand{\ua}{\uparrow}
\newcommand{\da}{\downarrow}

\newcommand{\dt}{{\text d}}
\newcommand{\etxt}{{\text e}}

\newcommand{\itxt}{{\text i}}

\newcommand{\ttxt}{{\text t}}

\newcommand{\Ktxt}{{\text K}}

\newcommand{\Ttxt}{{\text T}}

\newcommand{\Tx}{{\text T}}

\newcommand{\hh}{\hat{h}}

\newcommand{\ph}{\hat{p}}

\newcommand{\Hh}{\hat{H}}

\newcommand{\Abr}{\bar{A}}
\newcommand{\Bbr}{\bar{B}}

\newcommand{\Tc}{\mathcal{T}}

\newcommand{\pd}{\partial}
\newcommand{\rtarr}{\rightarrow}

\newcommand{\gt}{\tilde{g}}

\newcommand{\Deh}{\hat{\Delta}}

\newcommand{\pth}{\hat{\tilde{p}}}

\newcommand{\Ab}{{\bf A}}

\newcommand{\rb}{{\bf r}}

\newcommand{\At}{{\tilde{A}}}
\newcommand{\Bt}{{\tilde{B}}}

\newcommand{\dg}{\dagger}

\newcommand{\lan}{\langle}
\newcommand{\ran}{\rangle}

\newcommand{\al}{\alpha}
\newcommand{\be}{\beta}
\newcommand{\ga}{\gamma}

\newcommand{\de}{\delta}
\newcommand{\De}{\Delta}
\newcommand{\la}{\lambda}

\newcommand{\sig}{\sigma}

\newcommand{\e}{\epsilon}

\newcommand{\lt}{\left}
\newcommand{\rt}{\right}
\begin{document}
\title{Antiferromagnetic state in bilayer graphene}

\author{Maxim Kharitonov}

\affiliation{Center for Materials Theory, Department of Physics and Astronomy,
Rutgers University, Piscataway, NJ 08854, USA}
\date{\today}
\begin{abstract}

Motivated by the recent experiment of Velasco Jr. {\em et al.} [J. Velasco Jr. {\em et al.}, Nat. Nanotechnology 7, {\bf 156} (2012)],
we develop a mean-field theory of
the interaction-induced antiferromagnetic (AF) state in bilayer graphene at charge neutrality point at arbitrary perpendicular magnetic field $B$.
We demonstrate that the AF state can persist at all $B$.
At higher $B$, the state continuously crosses over to the AF phase
of the $\nu=0$ quantum Hall ferromagnet,
recently argued to be realized in the insulating $\nu=0$ state.
The mean-field quasiparticle gap is finite at $B=0$ and grows with increasing $B$, becoming quasi-linear
in the quantum Hall regime, in accord with the reported behavior of the transport gap.
By adjusting the two free parameters of the model, we obtain a simultaneous quantitative agreement between
the experimental and theoretical values of the key parameters of the gap dependence -- its zero-field value and slope at higher fields.
Our findings suggest that the insulating state observed in bilayer graphene in Ref.~\onlinecite{Velasco} is antiferromagnetic
(canted, once the Zeeman effect is taken into account) at all magnetic fields.

\end{abstract}
\maketitle

\begin{figure}
\includegraphics[width=.80\columnwidth]{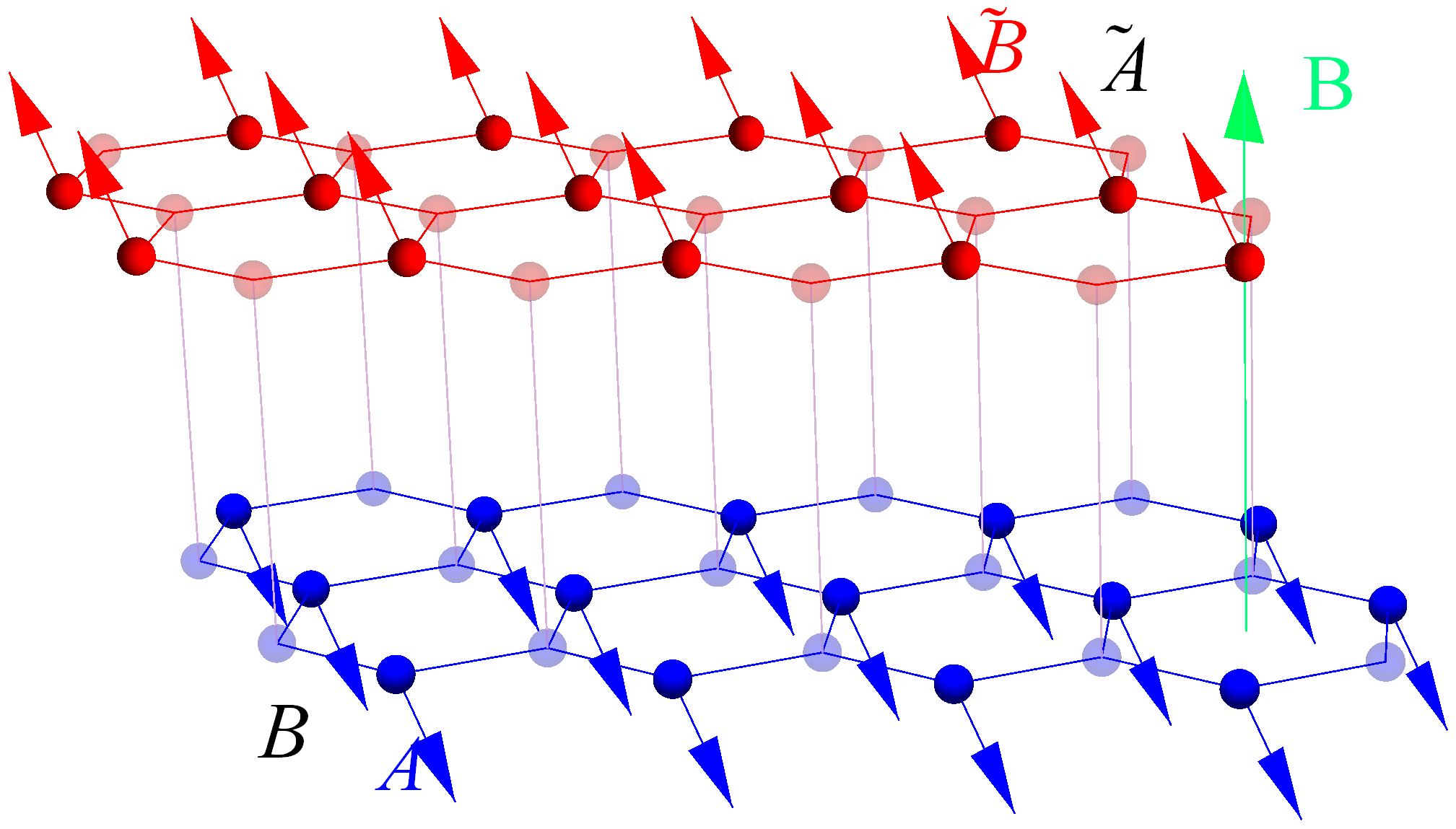}
\caption{
The antiferromagnetic (AF) state in bilayer graphene (BLG) at arbitrary orbital magnetic field.
If the Zeeman effect is neglected, as done in this paper for simplicity,
the $A$ and $\Bt$ sublattices, located in different layers, have arbitrary antiparallel spin polarizations, as shown.
The magnetization on the $B$ and $\At$ sublattices is negligible in the weak-coupling limit.
Once the Zeeman effect is included, the AF state transforms into the canted AF state~\cite{MK}, not shown here.
}
\label{fig:BLG}
\end{figure}

\section{Introduction}
Bilayer graphene (BLG)
presents an exciting arena for the observation of the correlated electron physics~\cite{
Velasco,Feldman,Zhao,Martin,Weitz,Mayorov,Freitag,
Min,VY,Zhang1,NL1,Lemonik,Vafek,NL2,Zhang2,Jung,Lemonik2,Cvetkovic,Zhu,
Barlas,APS,NL3,GorbarBLG,MK}.
The nearly quadratic dispersion of the electron spectrum about the charge neutrality point
makes the system susceptible to even weak interactions and at zero magnetic field,
allowing for instabilities towards various broken-symmetry phases.
A variety of correlated states at zero doping,
characterized by different ordering of the valley, layer, and spin degrees of freedom
have been predicted or considered~\cite{Min,VY,Zhang1,NL1,Lemonik,Vafek,NL2,Zhang2,Jung,Lemonik2,Cvetkovic,Zhu}.
At finite perpendicular magnetic field $B$,
quenching of the  kinetic energy facilitates the correlation effects.
In the quantum Hall (QH) regime,
the zero-density state transforms into the $\nu=0$
quantum Hall ferromagnet (QHFM)~\cite{Velasco,Feldman,Zhao,Martin,Weitz,Mayorov,Freitag,Barlas,APS,NL3,GorbarBLG,MK},
which also supports a number of interesting phases.

Recent transport experiments~\cite{Velasco,Weitz,Mayorov,Freitag} on high-quality suspended BLG samples provided compelling
evidence for the interaction-induced ground states both at $B=0$ and in the QH regime.
Several qualitatively different behaviors were reported.
In Refs.~\onlinecite{Weitz,Mayorov}, the zero-density state was insulating
in the QH regime (reached already at $B \gtrsim 1\Ttxt$),
showed metallic value of the two-terminal conductance $G \gtrsim e^2/h$ at $B=0$ and
a nonmonotonic behavior of $G$ at intermediate $B \lesssim 1\Ttxt$.
In Ref.~\onlinecite{Freitag},
in cleaner samples (labeled B2 therein), at $B=0$,
the differential conductance displayed signatures of the insulating gap with the minimal zero-bias conductance $G \approx 0.2 e^2/h$;
remarkably, at the same time, no fully developed insulating state was observed at higher $B$.
Finally, Ref.~\onlinecite{Velasco} reported a pronounced insulating state at all magnetic fields.
The transport gap was $E^\text{exp}_\text{gap} \approx 20 \Ktxt$ at $B=0$
and grew with increasing $B$, becoming linear in $B$ in the QH regime, with the slope $\dt E_\text{gap}^\text{exp}/\dt B \approx 5.5 \text{meV}/\Ttxt$,
as the state continuously crossed over to the $\nu=0$ QHFM state.

While all scenarios are equally interesting,
in this paper we concentrate on the theoretical description of the latter~\cite{Velasco} --
the insulating state at all magnetic fields.
We develop a mean-field (MF) theory of the insulating antiferromagnetic (AF) state in BLG, Fig.~\ref{fig:BLG},
at arbitrary perpendicular magnetic field.
We demonstrate that,
the AF phase can persist at all $B$,
continuously interpolating between
the earlier studied $B=0$~\cite{Min,Vafek,Zhang2,Jung}
and QH (AF phase of the $\nu=0$ QHFM)~\cite{MK} limits.
Most importantly, the obtained mean-field spectrum reproduces well the crucial experimental feature of Ref.~\onlinecite{Velasco}
-- the dependence of the transport gap on the magnetic field.
We obtain a simultaneous quantitative agreement between
the experimental and theoretical values of the key parameters of the gap dependence
-- its zero-field value and slope at higher fields --
by adjusting the two free parameters of the model.
Our findings further substantiate the conclusions about the AF phase in the QH regime~\cite{MK} and at $B=0$~\cite{Velasco},
suggesting the AF phase as the most likely candidate for the insulating state observed in Ref.~\onlinecite{Velasco}.

A large number of correlated phases
in BLG at $B=0$ predicted or considered
in theoretical literature~\cite{Min,VY,Zhang1,NL1,Lemonik,Vafek,NL2,Zhang2,Jung,Lemonik2,Cvetkovic,Zhu}
can be classified according to the properties of their charge excitations as
(i) bulk gapless (e.g., nematic~\cite{VY,Lemonik,Vafek,Lemonik2,Cvetkovic});
(ii) topologically nontrivial bulk gapped phases with gapless edge excitations
(e.g., quantum anomalous Hall (QAH)~\cite{NL2,NL3,Zhang2,Jung}, quantum spin Hall (QSH)~\cite{Zhang2,Jung,Lemonik2,Cvetkovic});
and (iii) fully gapped (bulk and edge) (e.g., ferroelectric~\cite{NL1,Jung,Lemonik2,Cvetkovic}, AF~\cite{Min,Vafek,Zhang2,Jung,Lemonik2,Cvetkovic}).
The AF phase was argued in Ref.~\cite{Velasco}
to be the most likely candidate for the insulating state at $B=0$,
while the phases of (i) and (ii) types can be ruled out with certain confidence,
since they should exhibit metallic two-terminal conductance $G \gtrsim e^2/h$.
The phases (i) or (ii) are more suitable candidates for the metallic low-field behavior
observed in Refs.~\cite{Weitz,Mayorov}.

In the QH regime,
the zero-density state transforms into the $\nu=0$ QHFM~\cite{Barlas,APS,NL3,GorbarBLG,MK}.
The generic phase diagram of the $\nu=0$ QHFM in BLG
was obtained in Ref.~\onlinecite{MK} and consists of four phases:
spin-polarized, antiferromagnetic
(canted, once the Zeeman effect is taken into account),
interlayer-coherent (at zero perpendicular electric field),
and fully layer-polarized.
Also, it was argued in Ref.~\onlinecite{MK} that the experimentally observed insulating $\nu=0$ QH state
in BLG is the AF phase of the $\nu=0$ QHFM.
This conclusion was reached by comparing the obtained phase diagram
with the experimental data of Ref.~\onlinecite{Weitz}
and was based on the argument that AF is the only phase
consistent with the observation of the insulator-insulator phase transitions
in the perpendicular electric field.
The same transitions are observed in Ref.~\onlinecite{Velasco} and thus
the same conclusion about the AF phase in the QH regime
can be made.

Crucially,
combined with the above conclusions,
the fact that the insulating state of Ref.~\onlinecite{Velasco}
shows a continuous crossover
between the zero-field and QH regimes
strongly suggests that the AF phase persists at all magnetic fields.
Here we theoretically demonstrate that this is indeed a feasible scenario.

\section{Model}
Our starting point is the Hamiltonian for
interacting electrons in the perpendicular magnetic field,
\begin{eqnarray}
    \Hh &= &\Hh_0+\Hh_\itxt, \nonumber\\
    \Hh_0 &=& \int \dt^2 \rb\, \psi^\dg \hh_0 \psi, \mbox{ }
    \hh_0 = \frac{1}{2 m}  \lt( \Tc_{z +} \pth_+^2  + \Tc_{z-} \pth_-^2 \rt),
\label{eq:H0} \\
    \Hh_\itxt &=& \frac{1}{2} \int \dt^2 \rb \, \sum\nolimits_{\al\be} \frac{4\pi}{m} g_{\al\be} :\!\![\psi^\dg \Tc_{\al\be} \psi]^2\!\!:.
\label{eq:Hi}
\end{eqnarray}
We describe electron dynamics in the framework of the two-band model~\cite{MF} of BLG, valid at energies
$\e \ll t_\perp$
below the interlayer hopping amplitude $t_\perp \approx 0.3 \text{eV}$.
At such energies, the wave-functions are predominantly localized on $A$ and $\Bt$ sublattices, located in different layers, Fig.~\ref{fig:BLG}.
The relevant degrees of freedom are joined into the  eight-component field operator
$\psi = (\psi_\ua,\psi_\da)^\ttxt$,
$\psi_\sig = (\psi_{KA}, \psi_{K \Bt}, \psi_{K'\Bt}, -\psi_{K'A})^\ttxt_{KK' \otimes \Abr\Bbr}$,
in the direct product $KK' \otimes \Abr \Bbr \otimes s$
of the valley, sublattice, and spin spaces, respectively.
We use the same basis as in Refs.~\onlinecite{Lemonik,Lemonik2}. Note that in this basis, the actual $A$ and $\Bt$ sublattices
are interchanged in the $K'$ valley;
therefore, to avoid confusion, we denote this sublattice space as $\Abr\Bbr$.
In Eq.~(\ref{eq:Hi}), $: \ldots :$ denotes normal ordering of operators and the summation goes over $\al,\be \in\{0,x,y,z\}$.
We set $\hbar=1$ everywhere in the paper, except for the quantum conductance value $e^2/h$.

In the kinetic energy term~(\ref{eq:H0}), $m$ is the effective mass,
$\pth_\pm =\pth_x \pm \itxt \pth_y$,  $\pth_\al= \ph_\al-\frac{e}{c} A_\al $,
$\ph_\al = -\itxt \pd_\al$ for $\al=x,y$, and $\text{rot}\, \Ab = (0,0,B)$.
In Eqs.~(\ref{eq:H0}) and (\ref{eq:Hi}) and below,  for $\al,\be,\ga \in \{0,x,y,z\}$,
\[
    \Tc_{\al\be\ga} = \tau^{KK'}_\al \otimes \tau^{\Abr \Bbr}_\be \otimes \tau_\ga^s,
    \mbox{ }
    \Tc_{\al\be} =\Tc_{\al\be 0},
    \mbox{ }
    \Tc_{z \pm} = \Tc_{z x} \pm \itxt \Tc_{z y},
\]
with the unity ($\tau_0 = \hat{1}$) and Pauli ($\tau_x$, $\tau_y$, $\tau_z$)
matrices in the corresponding subspaces.
To keep the analysis simpler,
we leave the orbital magnetic field as the only single-particle effect
and neglect the effects of warping and strain~\cite{MAF}:
the quite large extracted value $E_\text{gap}^\text{exp} \approx 20 \Ktxt$
of the transport gap suggests that the correlation effects dominate over these effects
under the experimental conditions of Ref.~\onlinecite{Velasco}.
We also neglect the Zeeman effect for the same reason:
for perpendicular field orientation, the actual canted AF phase~\cite{MK} should differ little from the AF phase.

Equation (\ref{eq:Hi}) is the most general form of the point two-particle interactions,
asymmetric in the $KK'\otimes \Abr \Bbr$ space,
allowed by the symmetry of the BLG  lattice~\cite{Lemonik,AKT,Lemonik2}.
The couplings satisfy the relations
$
    g_{\perp\perp} \equiv g_{xx} = g_{xy} = g_{yx} = g_{yy},
$
$
    g_{\perp z } \equiv g_{xz} = g_{yz},
$
$
    g_{z \perp } \equiv g_{zx} = g_{zy},
$
$
    g_{\perp 0 } \equiv g_{x0} = g_{y0},
$
$
    g_{0 \perp } \equiv  g_{0x} = g_{0y},
$
yielding  the total of nine independent couplings~\cite{AKT,Lemonik,Vafek,Lemonik2}.
The asymmetric channels [$(\al,\be)\neq(0,0)$]
arise from the Coulomb or electron-phonon interactions at the lattice scale
and may be assumed to have zero range.
For simplicity, we approximate the symmetric interactions [$(\al,\be)=(0,0)$] as local ones as well,
which is qualitatively justified for the following reasons.
First, in a typical experimental setup, a nearby metallic gate will screen the Coulomb interactions
beyond the distance $d$ from the BLG sample to the gate,
i.e., the interaction potential in the momentum space
is $V_0(q) = 2\pi e^2/(\kappa q)$ at $q d \gtrsim 1$ but $V_0(q) \sim e^2 d/\kappa$ at $q d \lesssim 1$
($\kappa$ is the dielectric constant of the environment).
Second, although the state studied below is gapped,
screening by the BLG electron system
is still efficient at momenta $q\gtrsim q_*\equiv \max(\sq{\De_z^0 m}, 1/l_B)$
above the scale set by either the correlation length $1/\sq{\De_z^0 m}$ of the gapped state ($\De_z^0$ is the zero-field gap defined below)
or the magnetic length $l_B=\sq{c/(eB)}$.
That is, the static polarization operator $\Pi(q \gtrsim q_*) \sim N m$ ($N=4$ is the discrete band degeneracy due to two valleys and two spin projections)
is finite at such scales, even though it vanishes in the long wavelength limit, $\Pi(q=0)=0$.
Since the physical quantities are determined by the screened potential
$
    V(q) = V_0(q)/[1+\Pi(q) V_0(q)]
$
in a finite range of momenta $q$ [no gauge-invariant quantity depends on $V(q=0)$ alone],
the Coulomb interactions are still well screened $V(q\gtrsim q_*) \approx 1/\Pi(q\gtrsim q_*)$
at relevant momenta $q\sim q_*$ and may thus be effectively represented
by a contact potential with the coupling constant $g_{00} \sim (\fr{m}{4\pi})/\Pi(q\sim q_*) \sim 1/N$.

There is no accurate knowledge of the coupling constants $g_{\al\be}$,
yet their ``bare'' values at the bandwidth $\sim t_\perp$ of the two-band model
determine the favored broken-symmetry ground state in BLG at zero doping.
At $B=0$, a systematic weak-coupling analysis of the many-body instabilities is carried out
within the RG approach~\cite{VY,Lemonik,Vafek,Lemonik2}.
In the QH regime, the interaction-induced $\nu=0$ state is studied within the framework of QHFMism~\cite{Barlas,APS,NL3,GorbarBLG,MK}.
Among the variety of predicted phases,
the AF phase was demonstrated to occur at both $B=0$ and in the QH regime,
under realistic assumptions about the values of the coupling constants $g_{\al\be}$.
We will now assume that the AF phase is the favored ground state both at $B=0$ and in the QH regime
and demonstrate that the AF phase then persists at all intermediate $B$
and that the two limits are adiabatically connected.

\section{Mean-field analysis}
We study the problem within the MF approach.
At $B=0$, the order parameter (OP)
$
    Q = \lan : \!\psi \psi^\dg \!: \ran
$
of the AF phase has the form
\beq
     Q=  Q_z^0 \, \tau_z^{KK'} \otimes \tau_z^{\Abr\Bbr}  \otimes \tau_z^s.
\label{eq:QB0}
\eeq
At finite magnetic field $B>0$,
due to the emergence of the $n=0,1$ Landau levels (LLs)~\cite{MF}
and the peculiar property of their wave functions to reside
on only one sublattice in each valley,
the OP {\em necessarily} acquires a component
$\tau_\da^{\Abr\Bbr} = \frac{1}{2} (1- \tau_z)^{\Abr\Bbr}$ in the $\Abr\Bbr$ space;
hence one needs to include the $\Tc_{z0z}$ component in the full OP of the AF phase.
The OP that describes the AF state at arbitrary magnetic field therefore has the form
\beq
    Q=  \tau_z^{KK'} \otimes (Q_0 \hat{1} + Q_z \tau_z)^{\Abr\Bbr}  \otimes \tau_z^s.
\label{eq:OP}
\eeq

\begin{figure}
\hspace{2mm}
\includegraphics[width=.85\columnwidth]{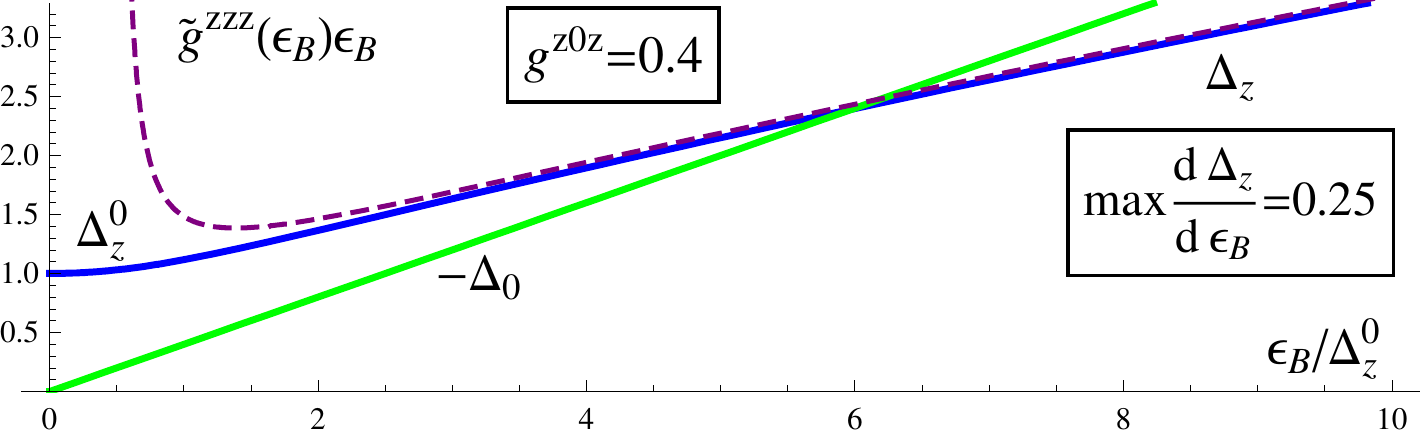}
\caption{The components $\De_0=-g^{z0z}\e_B$ and $\De_z$ of the mean-field potential (\ref{eq:De}) of the antiferromagnetic phase (AF) as functions of
$\e_B/\De_z^0$;
$\De_z$ is obtained by numerically solving the self-consistency equation, either Eq.~(\ref{eq:Deeq}) or (\ref{eq:Dezeq}).
The value $ g^{z0z}=0.4$ was used.
See caption to Fig.~\ref{fig:LLs} for details.
}
\label{fig:De}
\end{figure}

\begin{figure}
\includegraphics[width=.85\columnwidth]{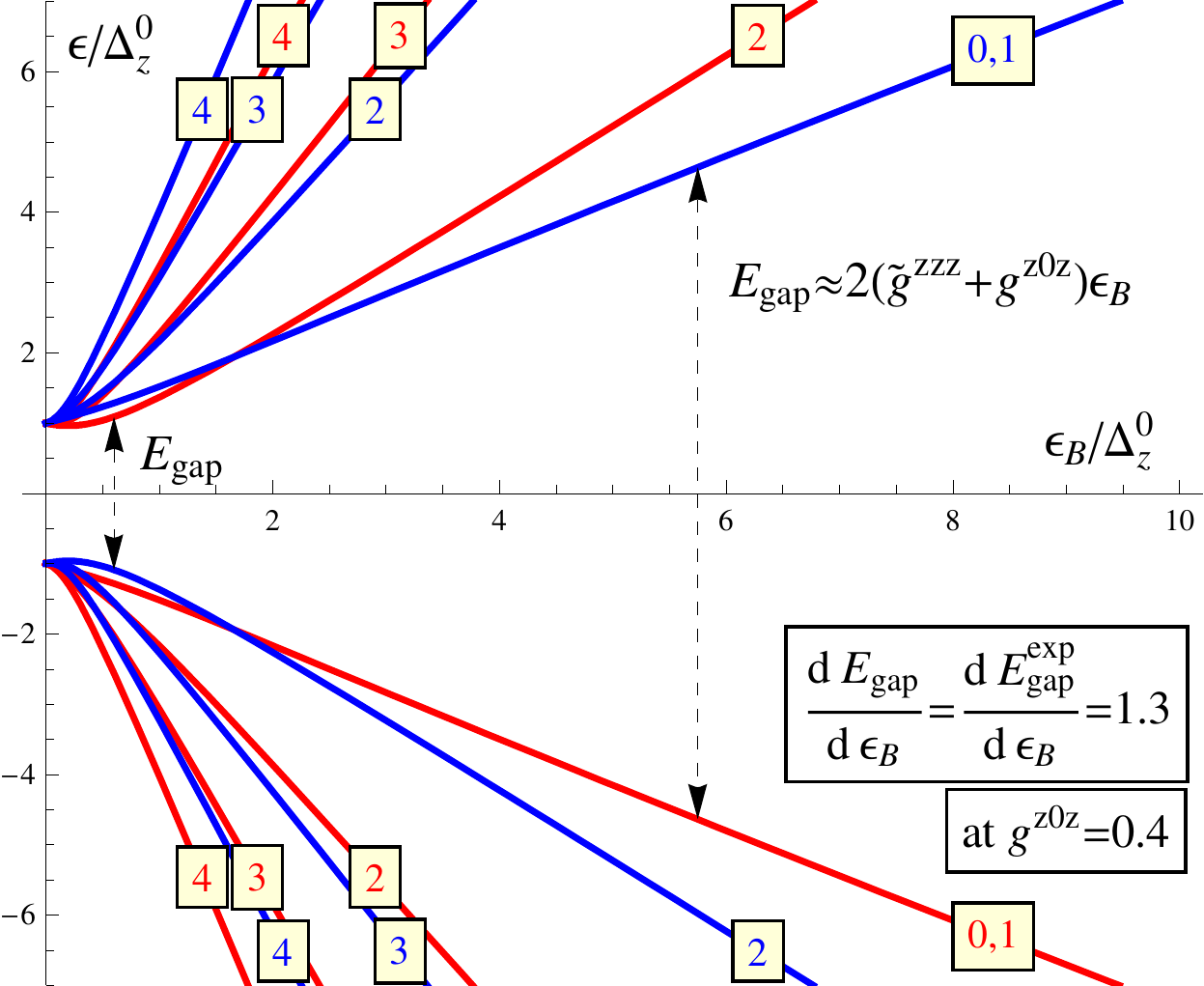}
\caption{Mean-field Landau level (LL) spectrum (\ref{eq:E01}),(\ref{eq:En}) of the antiferromagnetic (AF) phase
as a function of the magnetic field $B$,
obtained using the numerical solution for $\De_0$ and $\De_z$ shown in Fig.~\ref{fig:De};
the dependence on $B$ is expressed in terms of the ratio
$\e_B/\De_z^0$ of the cyclotron energy $\e_B= e B/(m c)$ and zero-field gap  $\De_z^0$.
Only the levels with $n\leq 4$ are shown (the numbers indicate $n$), $E_{n\pm K\ua}=E_{n\pm K'\da}$ (red), $E_{n\pm K\da}=E_{n\pm K'\ua}$ (blue).
The dependence of the gap $E_\text{gap}$ [Eq.~(\ref{eq:Egap})] between the positive and negative energy levels
on $\e_B$ closely reproduces that of the transport gap $E^\text{exp}_\text{gap}$ of the insulating state
observed in Ref.~\onlinecite{Velasco}, compare with Fig. 3 therein.
The used value $g^{z0z}=0.4$ provides quantitative agreement
with the experimental value $\dt E_\text{gap}^\text{exp} /\dt \e_B = 1.3$
(converted from $\dt E_\text{gap}^\text{exp} /\dt B = 5.5 \text{meV}/\Ttxt$ at $m=0.028 m_e$~\cite{Mayorov})
of Ref.~\onlinecite{Velasco} for the slope $\dt E_\text{gap}/\dt \e_B$ of the gap at higher fields, see Eqs.~(\ref{eq:DegapQH}), (\ref{eq:DegapQHslope}) and text.
}
\label{fig:LLs}
\end{figure}

Performing decoupling of interactions in Eq.~(\ref{eq:Hi}),
$
    \Hh_\itxt \rtarr \Hh_\text{i,mf} = \int \dt^2 \rb \, \psi^\dg \Deh \psi,
$
we obtain the MF potential
\beq
    \Deh = \tau_z^{KK'} \otimes (\De_0 \hat{1} + \De_z \tau_z)^{\Abr\Bbr}  \otimes \tau_z^s,
\label{eq:De}
\eeq
where
$\De_\al = - \frac{4\pi}{m} g^{z\al z} Q_\al$ ($\al=0,z$)
and
$g^{zzz} = g_{00} + g_{zz} + 4 g_{\perp\perp} - 2 g_{\perp z} - 2 g_{z\perp} -2g_{0\perp}+g_{0z}-2g_{\perp0}+g_{z0}$ and
$g^{z0z} = g_{00} + g_{zz} - 4 g_{\perp\perp} - 2 g_{\perp z} + 2 g_{z\perp} +2g_{0\perp}+g_{0z}-2g_{\perp0}+g_{z0}$.

Solving the eigenvalue problem for the Hamiltonian $\hh_0 + \Deh$,
we obtain the mean-field LL spectrum
\begin{eqnarray}
    E_{n\la\sig} = (\De_0 - \De_z) s_\la s_\sig , \mbox{ } n=0,1, \label{eq:E01}\\
    E_{n\pm\la\sig} = \De_0 s_\la s_\sig \pm \sqrt{\e_n^2 + \De_z^2}, \mbox{ } n \geq 2. \label{eq:En}
\end{eqnarray}
Here, $\e_n = \e_B \sqrt{n(n-1)}$ is the LL spectrum of the noninteracting BLG and $\e_B = 1/(m l_B^2)
\approx 1.3 \frac{m_e}{m} B[\Ttxt]\Ktxt$
is the cyclotron energy, with $m_e$ the electron mass.
Each state is characterized by the valley $\la=K,K'$ and spin $\sig=\ua,\da$ indices and $s_\la=\pm1$ and $s_\sig=\pm1$, respectively.

Calculating the OP (\ref{eq:OP}) in the eigenstate basis,
we obtain the self-consistency equations at zero temperature
\beq
    \De_z = g^{zzz}\e_B
        \lt( \sum_{n= 2}^{n_0} \frac{\De_z}{\sqrt{\e_n^2+\De_z^2}} + 1\rt),\mbox{ }
    \De_0 = - g^{z0z} \e_B.
\label{eq:Deeq}
\eeq
In the right-hand side of the equation for $\De_z$,
the unity represents the contribution from $n=0,1$ LLs,
while the sum from $n\geq2$ LLs.
We impose an ultraviolet energy cutoff $\e_0$ on the spectrum
and cut the otherwise logarithmically divergent sum by the large integer part $n_0=[\e_0/\e_B] \gg 1$.

Equations~(\ref{eq:OP})-(\ref{eq:Deeq})
are the key result of our work.
The solution of Eq.~(\ref{eq:Deeq}) for $\De_z$ and $\De_0$ determines the evolution of the order parameter (\ref{eq:OP})
of the AF state with the magnetic field and the MF quasiparticle spectrum (\ref{eq:E01}) and (\ref{eq:En}).
Below we discuss the key properties.

At $B=0$,  $\De_0 =0$ and the equation for $\De_z$ reduces to
\beq
    \De_z = g^{zzz} \int_0^{\e_0} \dt \e \frac{\De_z}{\sqrt{\e^2 + \De_z^2}}.
\label{eq:Dez0eq}
\eeq
Its solution $\De_z^0 = 2 \e_0 \exp(-1/g^{zzz})$ determines the gap in the spectrum at $B=0$.
The OP is given by Eq.~(\ref{eq:QB0}) with $Q_z^0= -\De_z^0 m/(4 \pi g^{zzz})$.

One can eliminate the cutoff $n_0$ from Eq.~(\ref{eq:Deeq}) for $\De_z$
using the standard procedure known from the BCS theory~\cite{AGD}.
Namely, one may represent the integral
\[
\int_0^{\e_0}  \frac{\dt \e}{\sqrt{\e^2 + \De^2_z}} =
        \sum_{n=2}^{n_0} f_n +f_0 + \ln \frac{\e_B}{\De_z} + o(1)
\]
that enters Eq.~(\ref{eq:Dez0eq}) for $\De_z^0$
by an {\em arbitrary} series with asymptotic
$
    f_0 + \sum_{n=2}^{n_0} f_n  = \ln(2 n_0) + o(1)
$
at $n_0=[\e_0/\e_B] \rtarr \infty$; the specific form of the series is a matter of convenience.
Adding and subtracting these two forms from the sum in Eq.~(\ref{eq:Deeq}),
one arrives at an {\em equivalent} equation
for $\De_z$,
\beq
    \frac{\e_B}{\De_z} = f_0 + \sum_{n=2}^{\infty} \lt( f_n \mm \frac{1}{\sqrt{n(n \mm 1) +(\De_z/\e_B)^2}} \rt) +  \ln \frac{\e_B}{\De_z^0}.
\label{eq:Dezeq}
\eeq
A slightly different mathematically, but equivalent
procedure of eliminating the high-energy cutoff from Eq.~(\ref{eq:Deeq}) was introduced earlier in Ref.~\onlinecite{TV}.

The form of Eq.~(\ref{eq:Dezeq})
shows explicitly 
that the functional dependence of $\De_z$ on $\e_B$ 
is, in fact, fully determined by one parameter, its value $\De_z^0$ at $B=0$.
The component $\De_0=-g^{z0z} \e_B$, 
in its turn,
is linear in $B$ and its slope is controlled by the coupling constant $g^{z0z}$.
The present theory is therefore described by {\em two} parameters,
the zero-field gap $\De_z^0$ and the coupling constant $g^{z0z}$.

In the QH regime $\e_B \gg \De_z^0$, it follows from Eq.~(\ref{eq:Dezeq}),
\beq
    \De_z= \gt^{zzz}(\e_B) \e_B, \mbox{ }
\tilde{g}^{zzz}(\e_B)= 1/[\ln(\e_B/\De_z^0) + C_0],
\label{eq:DezQH0}
\eeq
$C_0= \ln (2\etxt^{1-\ga}) + \sum_{n=2}^{\infty} [1/n- 1/\sqrt{n(n-1)}] \approx 0.674$.
We have introduced the notation $\gt^{zzz}(\e_B)$, since according to Eq.~(\ref{eq:Deeq}), Eq.~(\ref{eq:DezQH0})
can be interpreted as a QH-type dependence, akin $\De_0 = - g^{z0z} \e_B$,
with a ``renormalized'' coupling constant, $g^{zzz} \rtarr \gt^{zzz}(\e_B)=g^{zzz}/[1-g^{zzz}\ln(1.02 \e_0/\e_B)]$.

Thus, at higher fields, $\De_z$ is quasilinear in $\e_B$ with a logarithmically varying slope,
as observed earlier in Ref.~\onlinecite{TV}.
The OP equals
\beq
    Q = \frac{1}{2 \pi l_B^2} \tau_z^{KK'}  \otimes \tau_\da^{\Abr\Bbr}  \otimes \tau_z^s
    -
    \frac{1}{4 \pi l_B^2} \tau_z^{KK'}  \otimes \tau_z^{\Abr\Bbr}  \otimes \tau_z^s \frac{\de g^{zzz}}{g^{zzz}},
\label{eq:QQHFM}
\eeq
$\de g^{zzz} = \gt^{zzz}-g^{zzz}$.
The first contribution arises from $n=0,1$ LLs,
while the second one is the AF OP induced in the $n\geq 2$ LLs by LL mixing.
This OP describes the AF phase of the $\nu=0$ QHFM considered in Ref.~\onlinecite{MK}.

At intermediate fields, either Eq.~(\ref{eq:Deeq}) or (\ref{eq:Dezeq}) for $\De_z$ can be solved numerically.
The components $\De_0$ and $\De_z$ as functions of the magnetic field $B$ expressed in terms of $\e_B/\De_z^0$ are plotted in Fig.~\ref{fig:De}.
As anticipated, we find that upon applying the magnetic field the system preserves the AF order
and the AF state at $B=0$ [Eq.~(\ref{eq:QB0})] continuously crosses over to the AF phase of the $\nu=0$ QHFM [Eq.~(\ref{eq:QQHFM})].

The resulting mean-field LL spectrum (\ref{eq:E01}),(\ref{eq:En}) of the AF state is plotted in Fig.~\ref{fig:LLs}.
The MF quasiparticle gap
\beq
    E_\text{gap} = 2 \min(E_{2+K\ua},E_{0K\da})
\lbl{eq:Egap}
\eeq
is given by twice the energy of the lowest positive state.
The gap $E_\text{gap} = 2 \De_z^0$ is finite at $B=0$.

As seen from Figs.~\ref{fig:De} and \ref{fig:LLs},
in the major range of fields $\e_B \gtrsim \De_z^0$,
the gap is determined by $n=0,1$ LLs,
\beq
    E_\text{gap}/2 = E_{0K\da}= \De_z+|\De_0| \approx [\tilde{g}^{zzz}(\e_B) + g^{z0z}] \e_B,
\label{eq:DegapQH}
\eeq
and the formula (\ref{eq:DezQH0}) is accurate.
Thus, in the QH regime, reached already at $\e_B \gtrsim \De_z^0$, the gap (\ref{eq:DegapQH})
has a quasi-linear dependence on $\e_B$ associated with the QHFM physics;
the quasi-slope equals
\beq
 \frac{\dt E_\text{gap}}{\dt \e_B} \!=\!
  2\lt(\!
  \frac{\dt \De_z}{\dt \e_B} + g^{z0z}\!\rt),
\mbox{ }
    \frac{\dt \De_z}{\dt \e_B} = \tilde{g}^{zzz}(\e_B) - [\tilde{g}^{zzz}(\e_B)]^2.
\label{eq:DegapQHslope}
\eeq

At lower fields $\e_B \lesssim \De_z^0$, the gap $E_\text{gap}$
exhibit the following peculiar behavior, Fig.~\ref{fig:LLs}.
Since $\De_z $ grows quadratically at $\e_B/\De_z^0 \ll 1$,~\cite{TV}
whereas $\De_0$ is linear in $\e_B$,
the energies $E_{n+K\ua}=E_{n+K'\da}$ of $n\geq2$ LLs [Eq.~(\ref{eq:En})] initially decrease.
Thus at lower fields $\e_B \lesssim \De_z^0$,
the gap $E_\text{gap}=2 E_{2+K\ua}$ is determined by the $n=2$ LL and exhibits a nonmonotonic behavior in $\e_B$.
The $E_{2+K\ua}$ and $E_{0+K\da}$ LLs cross at a finite value of $\e_B\sim \De_z^0$,
above which the gap $E_\text{gap}=2 E_{0K\da}$ becomes determined by $n=0,1$ LLs, as discussed above.

These results for the gap dependence on the magnetic field were subsequently reproduced in Ref.~\onlinecite{TV}
within a slightly different approach.

\section{Comparison with experiment of  Ref.~1}

In the experiment of Ref.~\onlinecite{Velasco},
the transport gap $E_\text{gap}^\text{exp}(B)\equiv e V_\text{th}$ was defined
as the threshold value $V_\text{th}$ of the bias voltage $V$,
at which the low-temperature
$I(V)$-dependence experiences
a jump as the system switches from the insulating to the conducting state.
The obtained mean-field spectrum (Fig.~\ref{fig:LLs}) reproduces well the dependence of
the so-defined gap $E_\text{gap}^\text{exp}(B)$
on the magnetic field $B$, compare with Fig.~3 in Ref.~\onlinecite{Velasco}.
A distinct feature of $E_\text{gap}^\text{exp}(B)$
is that the linear dependence at higher $B$, if extrapolated to $B=0$, crosses the vertical axis at a value
only slightly below the actual zero-field gap. The calculated spectrum exhibits the same property.

A quantitative agreement between the key parameters of the experimental $E_\text{gap}^\text{exp}(B)$
and theoretical $E_\text{gap}(B)$ gap dependencies,
the zero-field value $ E_\text{gap}^\text{exp}(B=0)\approx 20\Ktxt$
and the slope
$
    \dt E^\text{exp}_\text{gap}/\dt\e_B= 1.3
$
at higher fields~\cite{conversion}, is achieved
by adjusting the two free parameters of the model, $\De_z^0$ and $g^{z0z}$.
First, the zero-field gap is fit by setting $\De_z^0 = E^\text{exp}_\text{gap}(B=0)/2\approx 10\Ktxt$.
Second, taking the typical slope  $\dt \De_z/\dt \e_B \approx 0.25 $
of the $\De_z(\e_B)$-dependence in the experimentally relevant range $1\Ttxt<B<4\Ttxt$ of fields,
we obtain from Eq.~(\ref{eq:DegapQHslope}) that the experimental slope is fit at $g^{z0z}\approx 0.4$.
The $\De_z(\e_B)$ and $\De_0(\e_B)$ dependencies in Fig.~\ref{fig:De}
and the spectrum in Fig.~\ref{fig:LLs} are presented for this value of $g^{z0z}$.

The predicted peculiar nonmonotonic behavior of $E_\text{gap}(B)$ at lower fields ($\e_B \sim \De_z^0$, $B\sim 1\Tx$)
is, however, not seen in the experimental $E_\text{gap}^\text{exp}(B)$-dependence of Ref.~\onlinecite{Velasco}.
Instead, $E_\text{gap}^\text{exp}(B)$ shows a monotonic growth with $B$. Several comments are in order in this regard.

On the theoretical side,
it is an open question whether the predicted low-field behavior
is a genuine feature of the AF state, rather than an artifact of the employed approximations,
and how robust it is to various perturbations.
As seen from Fig.~\ref{fig:LLs},
for the parameters that provide a fit of the slope at higher fields,
these features are numerically quite small
and could therefore be affected by several factors that were neglected in the present analysis,
such as the linear-in-momentum term in the spectrum or the Zeeman effect~\cite{Roy}.

On the experimental side,
first, the predicted features could simply
be beyond the resolution of the data of Ref.~\onlinecite{Velasco}.
Second, it is an open question whether the gap $E_\text{gap}^\text{exp}(B)$
defined as the bias threshold of the $I(V)$-dependence of a highly nonequilibrium state
accurately represents the true gap in the excitation spectrum of the ground state.
Measurements of the gap dependence on the magnetic field
via the thermally activated behavior of conductance in the linear response regime,
which would allow for a direct comparison with the present theory,
have not yet been reported.

For these reasons, we believe the low-field behavior deserves further investigation
before a conclusive comparison between the theory of the AF state and experiment in this regime can be made.

\section{Conclusion}
We developed a mean-field theory of the interaction-induced antiferromagnetic state in BLG at charge neutrality point at arbitrary
perpendicular magnetic field.
The theory reproduces well the key features of the recent experiment~\cite{Velasco} on suspended BLG samples:
persistence of the insulating state at all magnetic fields and the dependence of its transport gap  on the magnetic field.
At higher magnetic fields, the state crosses over to the antiferromagnetic phase of the $\nu=0$ QHFM, argued in Ref.~\onlinecite{MK}
to be realized in the insulating $\nu=0$ quantum Hall state.
The presented analysis suggests that the insulating state
observed in Ref.~\onlinecite{Velasco} is antiferromagnetic (canted, once the Zeeman effect is taken into account) at all magnetic fields.

\section*{Acknowledgements}
Author is thankful to O. Vafek, C. N. Lau, E. Andrei, and M. Foster for insightful discussions.
The work was supported by the U.S. DOE under Contract DE-FG02-99ER45790.


\begin{thebibliography}{99}
\bibitem{conversion}
The conversion of the slope to the units  $\dt E_\text{gap}^\text{exp}/\dt \e_B  \approx (\dt E_\text{gap}^\text{exp}/\dt B)/(1.3 m_e/m)\Ktxt/\Ttxt$
involves the value of the effective mass $m$.
We use the value $m \approx 0.028 m_e$ obtained in the experiment of Ref.~\onlinecite{Mayorov}.
\bibitem{Roy}
B. Roy, arXiv:1203.6340v1 (2012).
Note that the analysis performed therein, while taking the Zeeman effect into account, neglects the
component $Q_0 \Tc_{z0z}$ of the order parameter,
as well as several others that are generated once the Zeeman effect is present.
\end{thebibliography}
\end{document}